\documentclass[aps, pra, twocolumn, superscriptaddress, amsmath, longbibliography]{revtex4-1}
\usepackage{amssymb}
\usepackage{amsmath}
\usepackage{dcolumn}
\usepackage{graphicx}
\usepackage{mathrsfs}
\usepackage{appendix}
\usepackage{graphicx}
\usepackage{booktabs}
\usepackage{colortbl}

\definecolor{Dred}{RGB}{190,0,0}

\usepackage{url}
\usepackage[colorlinks]{hyperref}
\hypersetup{%
	plainpages=true,
	breaklinks=true,       
	hypertexnames=false,  
	pageanchor=true,
	colorlinks=true,
	linkcolor={blue},
	citecolor={red},
	urlcolor={blue},
	anchorcolor={black}
}

\def \hide#1{}

\begin{document}
\title{Chiral SQUID-metamaterial waveguide for circuit-QED}

\author{Xin Wang}
 \email{wangxin.phy@xjtu.edu.cn}
\affiliation{Institute of Theoretical Physics, School of Physics, Xi'an 
Jiaotong University, Xi'an 710049, People’s Republic of China,}

\author{Ya-Fen Lin}
\affiliation{Institute of Theoretical Physics, School of Physics, Xi'an 
Jiaotong University, Xi'an 710049, People’s Republic of China,}

\author{Jia-Qi Li}
\affiliation{Institute of Theoretical Physics, School of Physics, Xi'an 
Jiaotong University, Xi'an 710049, People’s Republic of China,}

\author{Wen-Xiao Liu}
\affiliation{Department of Physics and Electronics, North China University 
of \\
Water Resources and Electric Power, Zhengzhou 450046, People’s Republic of 
China}

\author{Hong-Rong Li}
\affiliation{Institute of Theoretical Physics, School of Physics, Xi'an 
Jiaotong University, Xi'an 710049, People’s Republic of China,}

\date{\today}

\begin{abstract}
Superconducting metamaterials, which are designed and fabricated with
structured fundamental circuit elements, have motivated recent developments of
exploring unconventional quantum phenomena in circuit quantum electrodynamics 
(circuit-QED). We propose a method to engineer 1D 
Josephson metamaterial as a 
chiral waveguide by considering a programmed spatiotemporal 
modulation on its effective impedance.
The modulation currents are in the form of travelling waves which phase velocities 
are much slower than the propagation speed of microwave photons. Due to the
Brillouin-scattering process, non-trivial spectrum regimes where photons can propagate unidirectionally
emerge. Considering superconducting qubits coupling with this 
metamaterial waveguide, we analyze both Markovian and non-Markovian quantum dynamics, 
and find that superconducting qubits can dissipate photons unidirectionally. Moreover, we show that our proposal can be extended a cascaded quantum network with 
multiple nodes, where chiral photon 
transport between remote qubits can be realized. Our work might open the 
possibilities to exploit SQUID metamaterials for realizing unidirectional 
photon transport in circuit-QED platforms.
\end{abstract}
\maketitle

\section{introduction}
The interaction between matter and quantized electromagnetic fields has been 
the central topic of quantum optics for more than half a 
century~\cite{Lamb1947,Scully1997,cohen1998atom,Walls07,Clerk10}. The 
impressive progresses in quantum electrodynamics (QED) and quantum 
technologies provide solid 
foundations for quantum information 
science~\cite{Clerk10,Buluta11,Xiang13,Underwood12,reiserer2015cavity,Chang2018}.
In past two decades, 
superconducting quantum circuits (SQCs) with Josephson 
junctions have been developed as a well-performed artificial platform for 
microwave photonics 
~\cite{nakamura1999coherent,Blais04,Koch07,rClarke08,You2011,Gu2017,
Ye2019,clerk2020hybrid,blais2021circuit,Gong2021}.
The 
versatility of SQCs stems from the flexibilities in both fabrication 
and controlling processes, which allows to 
realize many exotic quantum phenomena such as dynamical Casimir effects and 
ultrastrong light–matter 
couplings~\cite{Johansson09L,Johansson2010B,Wilson2011,Lhteenmki2013,Wilson2011,Beaudoin11,Kockum2019b,bourassa2009ultrastrong,Niemczyk2010}.

 
In circuit-QED the boson modes of microwave photons are 
routinely supported by conventional circuit elements such as LC resonators, 
transmission-line 
resonators and 1D open 
coplanar-waveguides~\cite{Majer2007,Goppl2008,Bourassa12,Clem2013,Doerner2018}.
Recently exploring intriguing QED phenomena with metamaterial structures in 
SQCs have attracted a lot of 
interests~\cite{Leib2012,Liu2017,Roushan2017a,Kollr2019,Ma2019,Carusotto2020,Mazhorin2022}.
Compared with standard circuit-QED elements, 
superconducting metamaterials can be designed and fabricated in various kinds 
of spatial structures, which might have exotic band spectrum and nontrivial 
vacuum modes~\cite{Mirhosseini2018,Scigliuzzo21}. For example, by 
engineering the hopping rates
among microwave resonators or SQC qubits, the photonic metamaterials analogue 
to 
topological lattices and strongly correlated matters were
realized~\cite{Roushan2017,Kim2021}. When the inductors and 
capacitors in the usual discrete 
model of a 1D transmission line are interchanged, the left-handed 
metamaterials can also be fabricated, which might work as a waveguide with 
an effective negative index of 
refraction~\cite{Caloz2004,Egger2013,Wang2019X,Messinger2019,Indrajeet2020}. 

The Josephson-chain metamaterials, which inherit the advantages of Josephson 
junctions such as 
nonlinearity and tunability, have been widely employed for 
different quantum engineering 
tasks~\cite{Masluk2012,Altimiras2013,weissl2014quantum,Weil2015,Krupko2018,Cosmic2018,Planat2020,Esposito2022,Sinha2022}.
For example, many-body quantum
optics in ultra-strong coupling regimes was successfully demonstrated in a 
Josephson chain platform by exploiting its high characteristic 
impedance~\cite{Martinez2019}. Moreover, given 
that each node junction is 
replaced with a superconducting quantum interference device (SQUID), the 
SQUID chain is possible to be engineered as a tunable high-impedance ohmic 
reservoir by applying position-dependent magnetic flux~\cite{Rastelli2018}. When the SQUID chain are biased with a space-time varying flux, 
analogue Hawking radiation can be
reproduced 
~\cite{Nation2009H,Nation2012s,Tian2017,Tian2019a,Lang2019,Blencowe2020}.
To create an event horizon, the group velocity of the 
modulation signals should be comparable to the propagation velocity. 

In contrast to proposals which are analogs of creating cosmological 
	particles~\cite{Nation2009H,Lang2019}, in this study we focus on the parameter regime 
	where 
	the velocity of the modulation signals is much slower than the photon's propagation speed, 
	which was rarely considered in previous studies.
The scenario is akin to realizing nonreciprocal sound propagation in an elastic 
metamaterial via spatiotemporally modulating the
stiffness~\cite{Swinteck2015,Trainiti2016,Yi2017,Cronne2017,Riva2019,Karkar2019,
Chen2019a,Attarzadeh2020}, where the Brillouin-scattering process will 
lead to chiral acoustic wave transport.
In this work we first derive 
the dispersion relation of the metamaterial waveguide by employing the generalized Bloch-Floquet expansion formula. We find that there are three 
non-trivial dispersion regimes, which respectively correspond to left/right chiral emission and band gaps. 
Then we show that the metamaterial waveguide can be engineered as a chiral quantum 
channel when coupling it to superconducting qubits.  The chiral transport is similar to the proposal of realizing nonreciprocal photonic transport via acoustic modulation~\cite{Calaj2019}. However, the modulating signal in our proposal 
is the electromagnetic current bias (rather than acoustic waves), which can be tuned much 
faster and freely. 
Our proposal can be employed for 
demonstrating chiral quantum optics widely discussed in 
Refs.~\cite{Cirac1997,Petersen2014,Lodahl2017,ZhangY2021,Guimond2020,Gheeraert2020,Solano2021,Kannan2022,Wang22A,Wang2022},
and provide a 
novel method to realize nonreciprocal transport of 
microwave photons without using the bulky ferrite 
circulator~\cite{Hogan1953,Caloz2018}.  Very 
recent studies with circuit-QED setups showed that chiral emission from qubit pairs (coupled 
to 
a 
common waveguide) can be realized via linear interference 
process~\cite{Guimond2020,Gheeraert2020,Solano2021,Kannan2022}. However, the chiral 
transport is restricted at the single-photon level. Our proposal does not have 
such limitations and can maintain the chiral behaviour even in the multi-excitation 
regime~\cite{Pichler2015,Mahmoodian2018,Prasad2020,Mahmoodian2020,Kusmierek22}.
In the last part, we show that our proposal can be extended 
as a chiral quantum 
network where remote nodes are mediated with no information back flow.

\section{Model}
\begin{figure}[tbph]
	\centering \includegraphics[width=8.8cm]{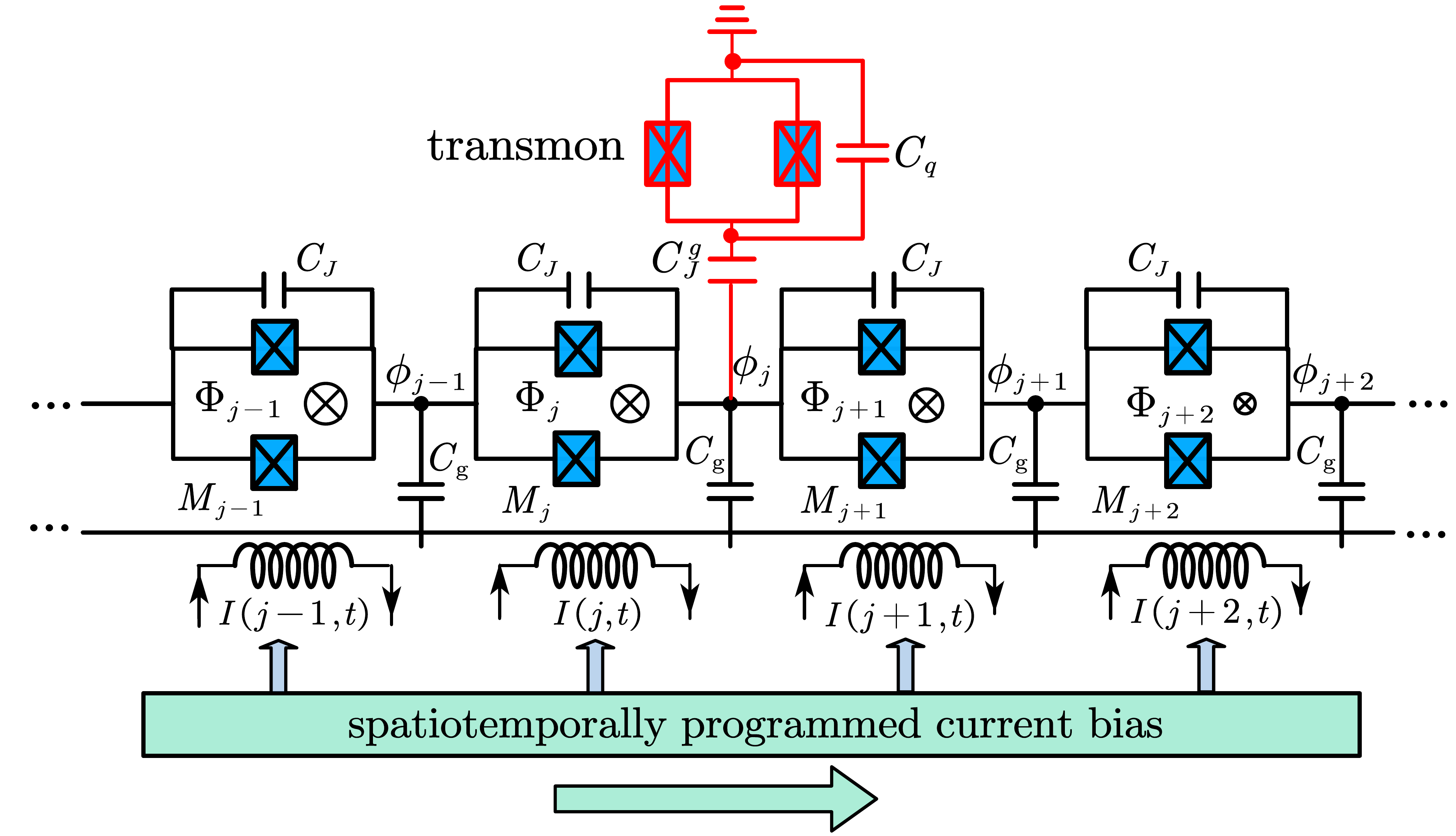}
	\caption{The proposed setup where a SQUID-metamaterial 
	waveguide 
	couples to a transmon qubit. The artificial waveguide is composed by a 
    SQUID array, which impedance is modulated by a programmed 
	current bias in each unit structure. The current signal $I(j,t)$ changing with 
	both SQUID's positions $j$ and time $t$ will spatiotemporally modulated 
	the flux 
	$\Phi_{j}$ through each SQUID via mutual inductance $M_j$. The 
	Josephson (ground) capacitance of each SQUID is denoted as $C_J$ ($C_0$). 
	The node flux $\phi_{j}$ at position $j$ follows the Kirchhoff current 
	relation in Eq.~(\ref{diff1}). A transmon 
	qubit, which is in the form of a split junction with Josephson capacitance 
	$C_q$, interacts with the waveguide at position $x_0=0$ via a coupling 
	capacitance $C_J^g$.}
	\label{fig1m}
\end{figure}

As shown in Fig~\ref{fig1m}, we consider that the Josephson-metamaterial waveguide  is composed by a SQUID array which allows microwave photons propagating along it~\cite{weissl2014quantum,Weil2015,Krupko2018}. Each node is
connected 
to the ground with a capacitance $C_{g}$. Two neighbor sites are separated with 
distance $d_0$. The SQUID can be viewed as a 
nonlinear Josephson inductance $L_{j}$ in 
parallel with a capacitance $C_{J}$. A series of current biases which can be 
programmed with external signals will produce site-dependent flux $\Phi_{j}$  
via mutual inductance $M_j$. The 
relation between $L_{j}$ and $\Phi_{j}$ is given 
by~\cite{Wallquist06,Sandberg2008,Eichler2014,Pogorzalek17,Eichler18,Wang2021}
\begin{equation}
	L_{j}=\frac{L_{0}}{\cos\left|\frac{\pi \Phi_{j}}{\Phi_{0}}\right|}, \quad 
	L_{0}=\frac{\Phi_{0}^{2}}{8\pi^{2}E_{s0}},
\end{equation}
where $E_{s0}$ is the Josephson energy of one junction. 
Defining the node flux as $\phi_{j}$, we obtain the following Kirchhoff 
current equation for the SQUID chain~\cite{weissl2014quantum,Weil2015}
\begin{eqnarray}
	&&\frac{\phi _{j+1}-\phi _j}{L_{j+1}}-\frac{\phi _j-\phi 
	_{j-1}}{L_{j}}=\notag \\
	&&C_g\ddot{\phi}_j+C_J\left( \ddot{\phi}_j-\ddot{\phi}_{j-1} \right)
	-C_{J}\left( \ddot{\phi}_{j+1}-\ddot{\phi}_j \right). 
	\label{diff1}
\end{eqnarray}
We assume that the current biases are composed of a static part 
and a time-dependent part, respectively. That is,  
\begin{equation}
\Phi_{j}(t)=M_{j}[I_{0}+\delta I f(j,t)],
\end{equation}
where $I_{0}$ ($\delta I$) is the amplitude of the DC (AC) part satisfying 
$\delta I f(j,t)\ll I_0$. We define $G_j(t)$ as 
\begin{eqnarray}
	G_j(t)=\frac{1}{L_{j}(t)}=\frac{1}{L_{0}}\left[\alpha_{0}+\delta\alpha 
	f(j,t)  \right],
\end{eqnarray} 
where 
\begin{gather}
\alpha_{0}=\cos\left(\frac{\pi M_{j}I_{0}}{\Phi_{0}}\right), \\
\delta\alpha=-\sin\left(\frac{\pi M_{j}I_{0}}{\Phi_{0}}\right)\frac{\pi M_{j}\delta I}{\Phi_{0}}.
\end{gather}

As discussed in Appendix A, given that $\phi _{j}$ varies slowly in the length 
scale $d_0$ (i.e., long wavelength limit), the difference 
Equation~(\ref{diff1}) can be written in a quasi-continuous form. 
Defining $c_{g}=C_g/d_0$ ($c_{J}=C_J/d_0$) as the ground (Josephson) capacitance per unit length,
the wave function in the quasi-continuous limit becomes~\cite{Wang2021} 
\begin{equation}
	c_{g}\frac{\partial^{2} \phi(x,t)}{\partial t^{2}}= c_{J}  
	d_0^2\frac{\partial^{4} \phi(x,t)}{\partial t^{2}\partial 
	x^{2}}+\frac{\partial}{\partial 
	x}\left[g(x,t)\frac{\partial\phi(x,t)}{\partial x}\right],
	\label{waveeq}
\end{equation}
where $g(x,t)$ is expressed as
\begin{equation}
	g(x,t)=\frac{1}{l_{J}}\left[\alpha_{0}+\delta\alpha f(x,t)\right], \quad l_{J}=\frac{L_0}{d_0}, 
\end{equation}
with $l_{J}$ being the Josephson inductance per unit length. The spatiotemporal modulation signal 
is encoded in $f(x,t)$. 
When $C_{J}=0$, Eq.~(\ref{waveeq})
is simplified as a transmission-line equation in which the impedance is modulated by a 
space-time-varying wave. 
A non-zero Josephson capacitance $c_{J}$ will entangle both spatial and temporal differentials 
together, which will produce a nonlinear dispersion for the SQUID waveguide.
In Appendix A and following discussions, we discuss its effects and give the conditions
where $c_J$ can be neglected.
\begin{table*}[tbp]
	{\normalsize \renewcommand\arraystretch{1.5}
		\begin{tabular}{>{\hfil}p{0.5in}<{\hfil}>{\hfil}p{0.5in}
				<{\hfil}>{\hfil}p{0.5in}<{\hfil}>{\hfil}p{0.5in}<{\hfil}>{\hfil}p{0.4in}<{\hfil}>{\hfil}p{0.5in}<{\hfil}>{\hfil}p{1.1in}<{\hfil}>{\hfil}p{1.1in}<{\hfil}>{\hfil}p{1.1in}<{\hfil}}
			\hline\hline  $d_{0}$ & $C_{g}$ & $C_{J}$ & $L_{0}$ & $\alpha$ & $\delta\alpha/\alpha$ 
			& $k_{d}/(2\pi)$ & $v_0$ & $v_d$   \\
			\hline $1~\mu \text{m}$ & 0.2~\text{fF} & 100~\text{fF} & 0.2~nH & 0.3  &$ 0.15$ & $0.25\times10^{4}\text{m}^{-1}$ & $2.7\times10^{6}~\text{m/s}$ & $(0.05\sim 0.08)v_0$ \\
			\hline\hline
	\end{tabular}}
	\caption{The parameters adopted for calculating the spectrum of the modulated SQUID-metamaterial waveguide. } \label{table1}
\end{table*}

The eigen-wave functions of the field are obtained by adopting a generalized Bloch-Floquet 
expansion~\cite{SLATER1958,Trainiti2016,Calaj2019}
\begin{equation}
	\phi_{lk}(x,t)=e^{i(\widetilde{\omega}_{l}(k) t-kx)}\sum_{n=-\infty}^{n=\infty}u_{ln}(k)e^{in(\Omega_d t-k_{d} x)},
	\label{phifs}
\end{equation}
where $\widetilde{\omega}_{l}$ is the quasi-energy of the $l$th band, and $k$ 
is the wave number. For convenience, we 
define
$v_0=1/\sqrt{l_J c_g}$ as the propagation velocity of the static SQUID 
metamaterial waveguide without modulation. Now we consider the bias current 
$g(x,t)$ varying periodically in both space and 
time, i.e.,  
$$g(x,t)=g(x+\lambda_d,t+T_d).$$ The simplest modulation signal is a travelling wave, and in 
this work we adopt 
$f(x,t)$ as
\begin{equation}
	f(x,t)=\cos{(\Omega_{d}t-k_{d}x)}.
	\label{fxt}
\end{equation}
The phase velocity is $v_d=\Omega_d/k_d$ with $k_d=2\pi/\lambda_d$ 
($\Omega_{d}=2\pi/T_d$) being the wave number (angular frequency). By 
substituting Eq.~(\ref{phifs}) into Eq.~(\ref{waveeq}), the 
dispersion relation is 
obtained by solving a quadratic eigenvalue problem~\cite{Trainiti2016}. 
Detailed derivations can be found in Appendix A.

According to the experiments in Refs.~\cite{Weil2015,Krupko2018,Martinez2019}, 
we list the parameters adopted for our numerical simulation in 
Table~\ref{table1}.  Although we can employ the 
generalized Floquet form in Eq.~(\ref{phifs}) to derive the spectrum, it does 
not mean that all kinds of modulation waves can 
produce stable eigenmodes in the SQUID transmission line. As 
discussed in Ref.~\cite{Cassedy1967}, 
when the modulation velocity $v_d$ is faster than $v_0$, the 
eigenfrequencies become complex, indicating that the oscillations are
time-growing and unstable. The high-speed modulation signal can be employed 
for conversing vacuum fluctuations into 
photons, which is analog to the Hawking effect in a gravitational 
system~\cite{Nation2009H,Blencowe2020}.

\begin{figure*}[tbph]
	\centering \includegraphics[width=15cm]{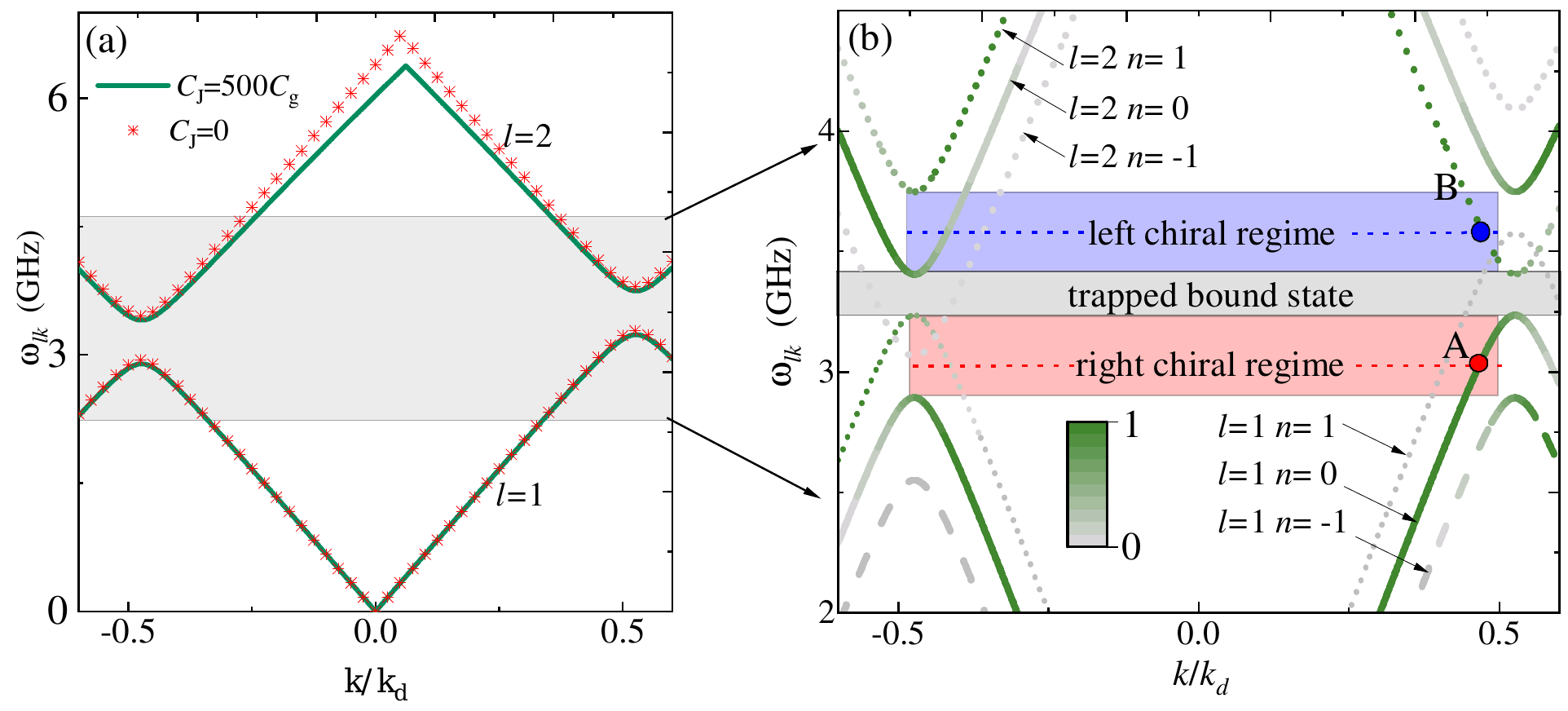}
	\caption{(a) Dispersion relation between quasi-energy $\omega_{l}(k)$ 
	and wave number $k$ for $C_J=500C_g$ and $C_J=0$, 
	respectively. The enlarged plot of the band gap [the area inside the box in 
	(a)] is depicted in (b). The dispersion relations for different Floquet 
	orders $\{l,n\}$ 
	[corresponding to time-dependent part in Eq.~(\ref{phifs})] are shown. 
    The distribution ratio of each Floquet order is 
	mapped with colors. The nontrivial spectra can be divided into three 
	parts. In the red (blue) area, the microwave photons propagate into the 
	right (left) direction; In the grey area, the microwave photon will be 
	localized due to no propagating modes. }
	\label{fig2m}
\end{figure*}

In this study, we focus on the parameter regimes with $v_d\ll v_0$. The first Brillouin zone (BZ) is limited within 
$k\in(-0.5k_d,0.5k_d]$.
There will be Brillouin-scattering
processes between modes $k$ and $k+k_d$ (which are wave numbers for the 
eigenmodes of the static waveguide) due to the conservation 
of momentum. Additionally, the conservation of energy also requires 
$\omega_l(k)\pm \Omega_d\simeq 
\omega_l(k+k_d)$. Since $\Omega_d$ is much smaller than $\omega_l$ at $k=k_d/2$, the modes around 
$k\simeq \pm 0.5k_d$ will interact with each other. The interactions between modes produces two band 
gaps, which is similar to the appearance of 
an anti-crossing point in a two-coupled-mode system.

We assume that the waveguide is long enough 
($L\rightarrow\infty$) to support the microwave photons propagating without 
reflection. By quantizing the field, the root-mean-square voltage operator $\hat{V}(x,t)$ is written 
as~\cite{Gu2017} 
\begin{equation}
	\hat{V}(x,t) \simeq -i \sum_{lk} \sqrt{\frac{\hbar \widetilde{\omega}_{l}(k)}{2C_t}} \left[ 
	\phi_{lk}^{*}(x,t) a_{lk}-\phi_{lk}(x,t) a^\dag_{lk} \right],
	\label{chargeQ}
\end{equation}
where $C_t = L c_g$ is the total capacitance of the waveguide, and $a_{lk}$ ($a_{lk}^{\dagger}$) is 
the annihilation (creation) operator for mode $k$ in the $l$th quasi-energy 
band. We numerically plot 
$\omega_{l}(k)$ changing with wave number $k$ in Fig.~\ref{fig2m}(a) by adopting parameters in Table~\ref{table1}. We find 
that an energy gap emerges between the two lowest bands in the first 
BZ $(-0.5k_d,0.5k_d]$. As 
discussed in Appendix A, under the following condition 
\begin{equation}
	\sqrt{\frac{c_J}{c_g}}\ll \frac{1}{\left( k+nk_{d} \right)d_0},
	\label{dis_con}
\end{equation}
the nonlinear effect led by $c_J$ can be neglected. 
By adopting the parameters in Table~\ref{table1}, one finds that 
$k_dd_0=0.25\times10^{-2}\ll 1$. According to Eq.~(\ref{dis_con}), even when 
$C_{J}=500C_{g}$, the nonlinear effect of Josephson capacitance is not 
apparent, and the dispersion relation is quite close to that with $C_{J}=0$. 

Equation~(\ref{phifs}) shows that the time-dependent part of $\phi_{lk}(x,t)$ 
depends on both 
band index $l$ and Floquet order
$n$, and the distribution ratio of the $n$th 
Floquet order is 
$|u_{ln}(k)|^2$ (note that $\sum_n |u_{ln}(k)|^2=1$) for a certain mode $lk$. 
Since the modulation signal $g(x,t)$ can be viewed as a
perturbation, $|u_{ln}(k)|$ decreases 
quickly with $n$ according to our numerical calculations. For the parameters in 
Table~\ref{table1},
it is accurate enough to
consider $n=0,\pm 1$, and the corresponding dispersion relation is plotted in 
Fig.~\ref{fig2m}(b) with
$|u_{ln}(k)|^2$ being mapped with colors.

The modulation signal $g(x,t)$ propagates unidirectionally (rather than a 
standing wave carrying opposite momentums $\pm k_d$), which will open two 
\emph{asymmetric} energy gaps located around $k=\pm 0.5 k_{d}$ (see Fig.~\ref{fig2m}). 
Consequently, the unconventional spectrum regime in Fig.~\ref{fig2m} can be 
divided into three parts. The gray area between two quasi-energy bands, where 
$|u_{ln}(k)|\simeq 0$ are valid for all Floquet orders $\{l,n\}$, is the 
conventional band gap with no 
propagating mode in the waveguide.  
In red (blue) area where point A (B) is located, the dispersion relation is 
asymmetric and the Floquet order $\{l=1,n=0\}$ 
($\{l=2,n=-1\}$) possesses the highest distribution ratio 
$|u_{ln}(k)|$. In these two regimes microwave photons propagate 
unidirectionally, which will be discussed below by considering a 
superconducting qubit interacting with 
this metamaterial waveguide.

\section{Chiral emission of superconducting qubits}
\subsection{Interaction Hamiltonian of metamaterial waveguide coupling to superconducting qubits}

As depicted in Fig.~\ref{fig1m}(a), we first consider the simplest QED setup where a  transmon qubit interacts with the metamaterial waveguide via 
a coupling capacitance $C^g_{J}$ at $x_0$. The transom qubit is composed by two 
identical junctions which form a SQUID loop, and its Hamiltonian is written 
as~\cite{Blais04,Koch07}
\begin{equation}
	H_{\text{q}}=4E_{C}(\hat{n}-n_{g})^{2}-2E^{q}_{J}\cos\!\left(\frac{\pi\Phi_{\text{q}}}{\Phi_{0}}\right)\cos{\hat{\phi}},
\end{equation}
where $\hat{n}$ ($\hat{\phi}$) is the charge (phase) operator of the transmon, 
$C_{\Sigma}=C^{q}_{J}+C_{q}$ with $C_{q}$ being
the Josephson capacitance, and $E^{q}_{J}$ ($E_{C}=e^{2}/(2C_{\Sigma})$) is the 
Josephson (charging) energy of the 
junction. The interaction Hamiltonian 
between the transmon and the waveguide is written as~\cite{Koch07} 
\begin{equation}
H_{\text{int}}=2e\frac{C^{q}_{J}}{C_{\Sigma}}\hat{V}(x,t)\hat{n}.
\end{equation}
In the limit $E_{C}\ll E_{J}^q$, the transmon can be viewed as a Duffing 
nonlinear oscillator. Given that only the two lowest energy levels are considered, 
$H_{\text{q}}$ can be approximately written as~\cite{Koch07}
\begin{equation}
H_{\text{q}}=\frac{1}{2}\omega_q\sigma_z, \qquad 
\omega_q=4\sqrt{E_CE^{q}_{J}}-E_C,
\end{equation}
where the charge operator is
\begin{equation}
\hat{n}=-i\sqrt[4]{\frac{E_{J}^q}{4E_{C}}}\frac{\sigma_{-}-\sigma_{+}}{\sqrt{2}}.
\end{equation}
In the rotating frame of $\omega_q\sigma_z/2+\sum_{lk}\hbar 
\widetilde{\omega}_{l}(k)a^\dag_{lk}a_{lk}$
and by adopting the rotating-wave 
approximation, we rewrite $H_{\text{int}}$ as
\begin{eqnarray}
	H_{\text{int}}&=&\hbar g_{0}\sum_{lk}e^{i\omega_{q}t}\sigma_{+}a_{lk}\phi^{*}_{lk}(x_c,t)+\text{H.c.}\notag \\
	&=&\hbar \sum_{lk} g_{lk}\sigma_{+}a_{lk}e^{i\omega_{q}t}  \notag \\
	&\times &
	\left[e^{-i\widetilde{\omega}_{l}(k) 
	t}\sum_{n=-\infty}^{n=\infty}u^*_{ln}(k)e^{-in\Omega_d t}\right]+\text{H.c.},
	\label{Hinto}
\end{eqnarray} 
where the coupling position is set at $x_0=0$ without loss of generality, and 
$g_{lk}$ is the coupling strength which is derived as 
\begin{equation}
g_{lk}=\sqrt{2}\frac{eC^{q}_{J}}{\hbar C_{\Sigma}}\sqrt[4]{\frac{E_{J}^q}{4E_{C}}}\sqrt{\frac{\hbar \widetilde{\omega}_{l}(k)}{2C_t}}. 
\label{glk}
\end{equation}
Here we take the transmon as an example. As discussed in experimental work in 
Refs.~\cite{Rastelli2018,Martinez2019,Wang2021}, the SQUID-metamaterial 
waveguide can also interact with a flux or charge qubit, and all these 
circuit-QED 
setups can be employed to demonstrate the unconventional emission behaviors 
discussed in this work.

\subsection{Chiral emission in Markovian regime}
\begin{figure*}[tbph]
	\centering \includegraphics[width=16cm]{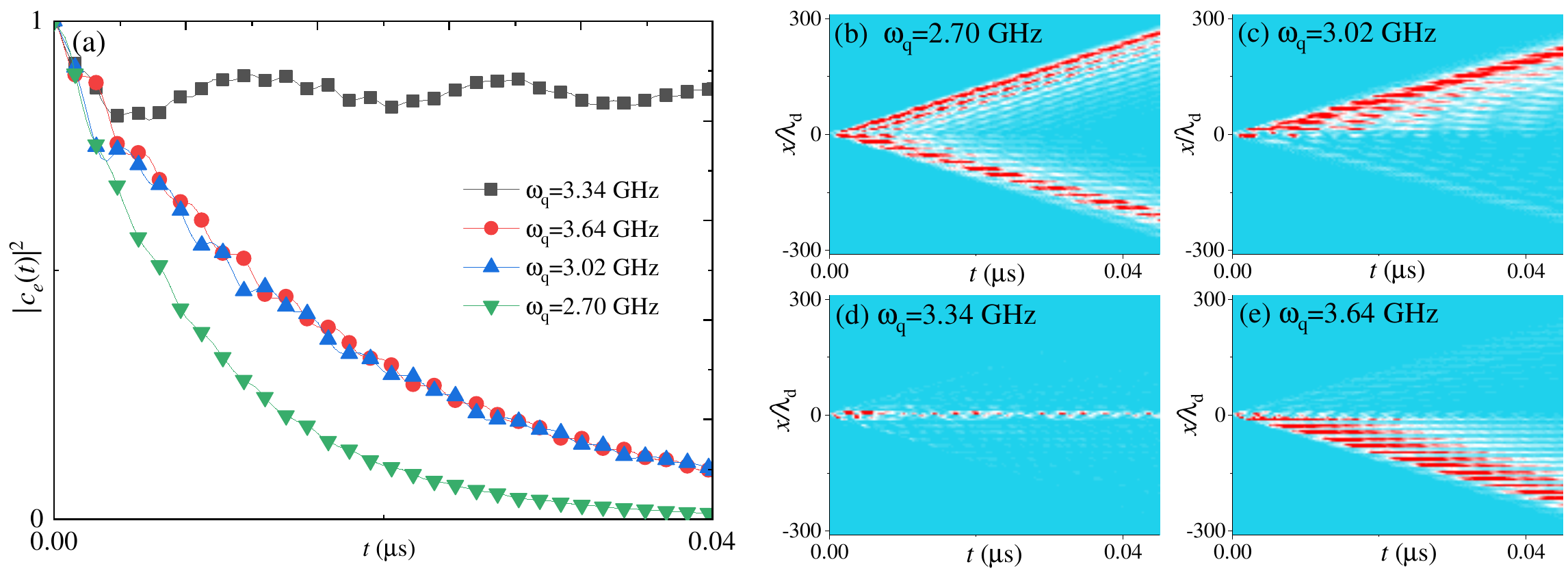}
	\caption{(a) The time-dependent evolution $|c_e(t)|^2$ for different 
	transmon frequencies, and the field 
	distributions are shown in (b-e), respectively. The corresponding relations 
	are (b) bidirectional  
	emission $\rightarrow\omega_q=2.70~\text{GHz}$, (c) right 
	emission $\rightarrow\omega_q=3.02~\text{GHz}$, (d) trapped bound state 
	$\rightarrow\omega_q=3.34~\text{GHz}$ and (e) left
	emission $\rightarrow\omega_q=3.64~\text{GHz}$. The system's parameters are 
	adopted from Table~\ref{table1} with $v_d=0.05v_0$.}
	\label{fig3m}
\end{figure*}
Similar to the standard spontaneous emission process, the intensity of 
radiation field should be narrowly centered around the atomic 
transition frequency $\omega_{q}$. Therefore, by replacing 
$\widetilde{\omega}_{l}(k)$ in Eq.~(\ref{glk}) with $\omega_{q}$, $g_{lk}$ becomes 
mode-independent, i.e., $g_{lk}\simeq g_0$. We assume that a single excitation is initially in the 
transmon qubit, while the waveguide is in vacuum state $|0\rangle$. Due to the 
conservation of excitation number for the Hamiltonian in Eq.~(\ref{Hinto}), the system's state at time $t$ can be expressed 
in the single-excitation subspace as
\begin{equation}
|\Psi(t)\rangle=c_{e}(t)|e,0\rangle+\sum_{l,k}c_{lk}(t)|g,1_{lk}\rangle,
\label{states}
\end{equation}
where $|1_{lk}\rangle$ represents a single photon being excited in mode 
$a_{lk}$. 
The evolution governed by $H_{\text{int}}$ can be derived from the following coupled differential equations
\begin{widetext}
\begin{eqnarray}
	\frac{\partial c_{e}(t)}{\partial t}&=&-ig_{0}\sum_{lkn}c_{lk}  
	u_{ln}(k)e^{[i\Delta_{l}(k)-in\Omega_d] t}, \label{cet} \\
	\frac{\partial c_{lk}(t)}{\partial t}&=&-ig_{0}c_{e}(t)e^{-i\Delta_{l}(k)t}  \sum_{n}u_{ln}^{*}(k)e^{in\Omega_d t}, \label{clkt}
\end{eqnarray}
where $\Delta_{l}(k)=\omega_{q}-\widetilde{\omega}_{l}(k)$ is the frequency 
detuning between the qubit and the mode $a_{lk}$. By substituting the integration 
form of Eq.~(\ref{clkt}) into Eq.~(\ref{cet}), we obtain
\begin{eqnarray}
	\frac{\partial c_{e}(t)}{\partial t}&=&-g_{0}^{2}\int_{0}^{t}dt'G(t,t') c_{e}(t'), \\
	G(t,t')&=&\sum_{lknn'}e^{-i(n-n')\Omega_d t} u_{ln}(k) u_{ln'}^{*}(k)e^{i[\Delta_{l}(k)-n\Omega_d](t-t')},   \label{decay1}
\end{eqnarray}
\end{widetext}
where $G(t,t')$ is the time-delay correlation function. Given that $n\neq n'$,
$e^{-i(n-n')\Omega_d t}$ are fast oscillating terms, and their contributions to 
the evolution will be significantly suppressed when the decaying time scale is 
much longer than the oscillating period $\Omega_d^{-1}$. Under these 
conditions, 
we can only keep the resonant term $n=n'$, which is similar to the 
rotating-wave approximation~\cite{Calaj2019}. Finally we obtain 
\begin{eqnarray}
	G(t,t')&=&\sum_{lkn}|u_{ln}(k)|^2 e^{i[\Delta_{l}(k)-n\Omega_d](t-t')} 
	\notag \\
	&=&\sum_{ln}\frac{L}{2\pi}\int dk
	|u_{ln}(k)|^{2}e^{i[\Delta_{l}(k)-n\Omega_d](t-t')}. 
\end{eqnarray}
In the emission spectrum, the intensity of the field will center at the modes 
satisfying the resonant condition
\begin{equation}
	\Delta_{l}(k)-n\Omega_d=0,
	\label{rescon}
\end{equation} 
which solutions of $k$ are denoted as $k_{ln}$. For example, in 
Fig.~\ref{fig2m}(b) for the frequency at the dashed position in the red regime, we 
mark the resonant positions (green dot $A$) for $\{l=1,n=0\}$.
Around each $k_{ln}$, one can approximately derive the dispersion relation as
\begin{equation}
	\Delta_{l}(k)=\Delta_{l}(k_{ln})+v_{ln}(k-k_{ln}), 
\end{equation}
where
$$v_{ln}=\frac{\partial \Delta_{l}(k_{ln})}{\partial k}\Big|_{k_{ln}}$$ 
is the group velocity. When the qubit transition frequency is far away from the 
band edges, $G(t,t')$ can be written as
\begin{eqnarray}
	G(t,t')&\simeq& \sum_{ln}\frac{L}{2\pi}|u_{ln}(k_{ln})|^{2}\int_{-\infty}^{\infty}d(\delta k)
	e^{-iv_{ln}\delta k(t-t')} \notag\\
	&=&\sum_{ln}\frac{L|u_{ln}(k_{ln})|^{2}}{|v_{ln}|}\delta(t-t').
	\label{Gttf}
\end{eqnarray}
The delta function in Eq.~(\ref{Gttf}) is valid given that the 
bandwidth of the waveguide's spectrum is approximately infinite compared with 
the interaction strength. By substituting Eq.~(\ref{Gttf}) into 
Eq.~(\ref{decay1}), we obtain
\begin{eqnarray}
	\frac{\partial c_{e}(t)}{\partial t}&=&-\frac{\Gamma_{T}}{2} c_{e}(t), \quad \Gamma_{T} =(\Gamma_{R}+\Gamma_{L}),\\
	\Gamma_{R(L)}&=&\Gamma_0\sum_{ln}|u_{ln}(k_{ln})|^{2}\Theta(\pm v_{ln}), 
	\label{GammaRL}
\end{eqnarray}
where $\Theta(x)$ is the
Heaviside step function, $\Gamma_{R(L)}$ corresponds to the decay rate to
the right 
(left) hand of the waveguide, and $\Gamma_{0}$ is the characteristic decay 
rate for the qubit which is derived as 
\begin{equation}
	\Gamma_{0}=\frac{g_{0}^{2}L}{ v_0}=\frac{1}{\hbar 
	v_J}\left(\frac{eC^{q}_{J}}{ 
	C_{\Sigma}}\right)^2\sqrt[2]{\frac{E_{J}^q}{4E_{C}}}\frac{ 
	\omega_{q}}{2c_g}.
	\label{gamma0}
\end{equation} 

In the single-excitation subspace expressed in Eq.~(\ref{states}), the 
photonic 
wave function in real space is~\cite{Scully1997}
\begin{eqnarray}
	\psi_{\gamma}(x,t)=\sum_{lkn}c_{lk}(t)e^{ikx}u^{*}_{ln}(k)e^{-in(\Omega_d 
	t-k_{d} x)}.
	\label{field_D}
\end{eqnarray}
The photonic energy decaying into the right (left) hand side is expressed as
\begin{equation}
	\Phi_{R/L} = \left| \int_{0}^{\pm \infty} |\psi_{\gamma}(x',t)|^2 dx' \right|, \quad t\rightarrow\infty.
	\label{field_dis}
\end{equation}
By neglecting the local decoherence and dephasing of the transmon, the 
chiral factor is defined as~\cite{Lodahl2017} 
\begin{equation}
	\beta_{\pm}=\frac{\Gamma_{R(L)}}{\Gamma_{R}+\Gamma_{L}}=\frac{\Phi_{R(L)}}{\Phi_{R}+\Phi_{L}}.
	\label{beta_ana}
\end{equation}

As indicated in Eq.~(\ref{GammaRL}), the directional emission rates depend on 
both distributions ratios $|u_{ln}(k_{ln})|$ and 
group velocities 
$v_{ln}$. For example, given that the qubit frequency lies in the red regime of 
Fig.~\ref{fig2m}(b), the Floquet order 
$\{l=1,n=0\}$ has the largest 
$|u_{ln}(k_{ln})|$ (see point A), and the corresponding $v_{ln}$ is positive. 
Consequently, the photon will be chirally emitted to the right direction. In 
the blue regime (point B), the emission chirality will be reversed.
Below we will numerically discuss the emission behaviors in different 
frequency regimes.

\subsection{Numerical discussions}
The chiral spontaneous emission process can well be described by the master 
equation, which however discards much
information due to a lot of approximations. For example, master equation cannot 
describe the directional field distribution and Non-Markovian dynamics led by  
band-edge effects. To avoid those problems, we numerically simulate the unitary 
evolution governed by Hamiltonian in 
Eq.~(\ref{Hinto}) by discretising the modes of the modulated waveguide in the 
momentum space. The photonic field in the waveguide is 
recovered from 
Eq.~(\ref{field_D}).  The details about numerical methods are
presented in Appendix B. 

By changing the transmon's frequency, we plot the evolutions 
of $|c_e(t)|^2$ in Fig.~\ref{fig3m}(a). When $\omega_q$ is in the conventional 
dispersion regime [$\omega_q/(2\pi)=2.70~\text{GHz}$], the transmon 
exponentially decays its energy, 
and the photonic field is symmetrically emitted into both left and right side [see 
Fig.~\ref{fig3m}(b)]. In the right chiral regime with
$\omega_q/(2\pi)=3.02~\text{GHz}$ [see red horizontal line in Fig.~\ref{fig2m}(b)], the 
solutions 
for Eq.~(\ref{rescon}) correspond to the intersection points with the dispersion curves of 
different Floquet orders. The intersection point A with the Floquet order $\{l=1,n=0\}$ has 
the 
largest distribution ratio $|u_{ln}(k)|\simeq 1$, while the other $|u_{ln}(k)|$ are of 
extremely low amplitudes. The 
group velocity for point A is positive. Therefore, the 
transmon will chirally emit photons 
into the right part of the waveguide, as indicated by Eq.~(\ref{GammaRL}). Similarly, when 
transmon frequency 
is in the blue regime, most of the emitted photonic field will distribute on its left hand 
side. The chiral field distributions changing 
with time are shown in 
Fig.~\ref{fig3m}(c, e), 
respectively. Moreover, the chiral 
factor is about $\beta_{\pm}\simeq 0.95$, indicating that this 
SQUID-metamaterial waveguide can be implemented as a well-performance 
directional quantum bus. 
\begin{figure}[tbph]
	\centering \includegraphics[width=8.8cm]{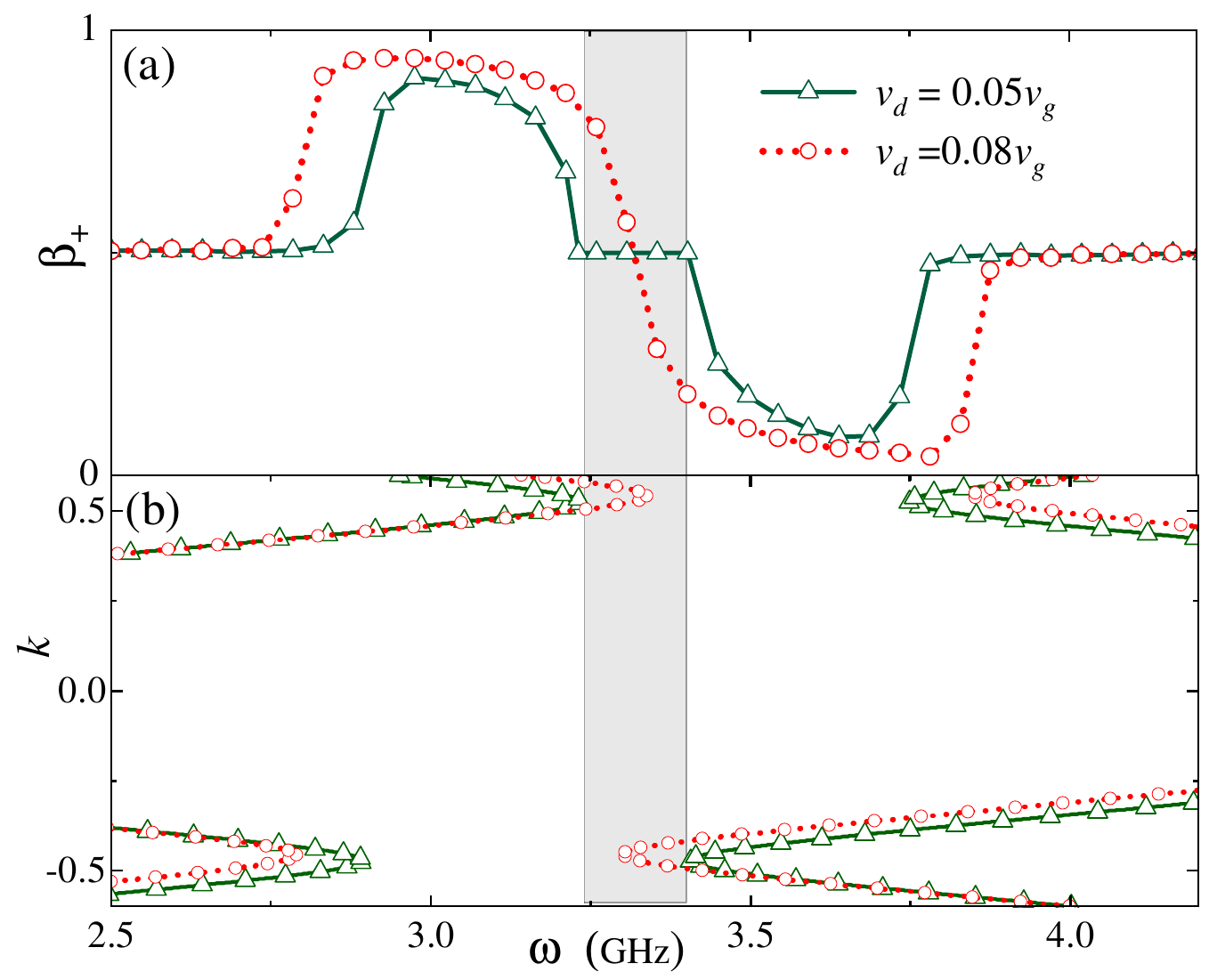}
	\caption{(a) The chiral factor $\beta_+$ changes with
	$\omega_q$ by setting modulation group velocities as $v_d=0.05v_g$ and 
	$v_d=0.08v_g$. (b) The corresponding dispersion relation of the 
	quasi-energy $\omega_{l}(k)$ of 
	the 
	waveguide.}
	\label{fig4m}
\end{figure}

When the qubit frequency lies within the band gap [the gray regime in 
Fig.~\ref{fig2m}(b)], the distribution 
ratios for all Floquet orders are around zero, i.e., $|u_{ln}(k)|\simeq 0$, indicating
that there is no resonant mode which can lead to exponential decay of the transmon 
qubit.
In this scenario,
the transmon only decays its partial energy 
into the waveguide [see Fig.~\ref{fig3m}(a) for
$\omega_q/(2\pi)=3.34~\text{GHz}$], and the field is localized 
around the coupling position without propagating outside, as shown
Fig.~\ref{fig3m}(d)]. This 
unconventional evolution can be understood from Fig.~\ref{fig2m}(b): in the 
gray 
regime, the coefficients of all the Floquet 
orders are around zero, i.e., 
$|u_{ln}(k_{ln})|\simeq 0$. Due to no resonant mode, only the modes which are 
of large detuning will 
contribute 
significantly to the dynamics. Those modes are of large density due to band 
edge effects, and the 
field will be localized in the form of 
bound state. The scenario is akin to an emitter being prevented from spontaneous emission when 
it is trapped by the band gap of a photonic crystal 
waveguide~\cite{Goban2014,GonzlezTudela2015,Douglas2016}.

In this proposal the modulation signal is programmable. Therefore, the 
dispersion relation of the waveguide can be tailed freely, which 
enables us to control the qubit dynamics on demand.
For example, given that the modulation phase velocity $v_d$ is switched 
oppositely, the 
chiral direction with a certain qubit frequency is also reversed. 
In Fig.~\ref{fig4m}(a), we plot the chiral factor 
$\beta_+$ 
changing with transmon frequency by setting modulation velocity as 
$v_d=0.05v_g$ and 
$v_d=0.08v_g$, respectively. When $v_d=0.05v_g$, 
the quasi-energy band $l=1$ and $l=2$ are separated by 
a finite band gap [see gray area in Fig.~\ref{fig2m}(b)]. The gap leads to the trapped
bound state when the qubit 
frequency lies within this regime [see Fig.~\ref{fig3m}(d)]. As discussed in 
Ref.~\cite{Trainiti2016}, when increasing the 
modulation velocity, the gap disappears, which can be found from the dispersion relation 
for $v_d=0.08v_g$ depicted in Fig.~\ref{fig4m}(b). The chirality will be 
smoothly switched from left to right without any gaps when increasing the qubit frequency. 
With a larger $v_d$ the chirality is
enhanced and the directional bandwidth becomes wider [see Fig.~\ref{fig4m}(a)], which is due 
to that the 
Brillouin-scattering process emerges between the modes with a large energy difference. 
However, 
as discussed in 
Refs.~\cite{Trainiti2016,Cassedy1967}, 
when the modulation velocity $v_d$ is comparable to $v_0$, the waveguide's eigenfrequencies
become complex, indicating the field is time-growing and unstable. Due 
to 
this, the modulation velocity $v_d$ should be much smaller than $v_0$ to avoid the unstable 
phenomena emerging in the whole setup.


\section{Chiral photon flow between two quantum nodes}
\begin{figure}[tbph]
	\centering \includegraphics[width=8.8cm]{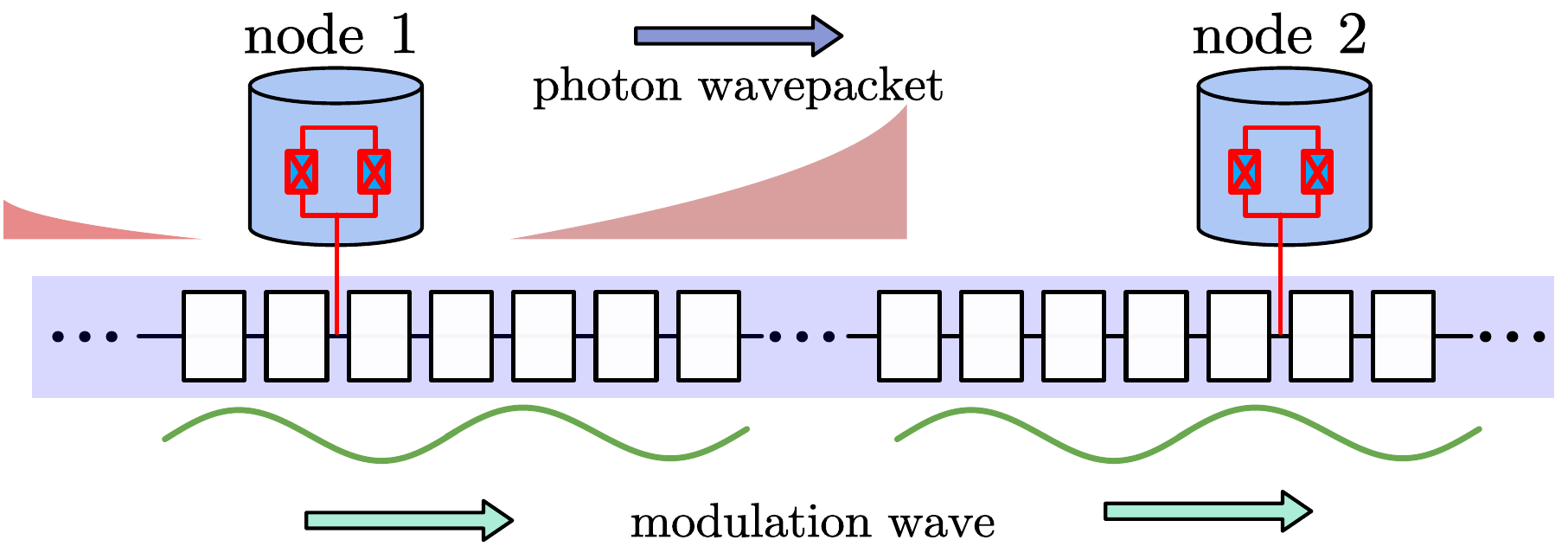}
	\caption{By mediating remote nodes, the metamaterial 
	waveguide can work as a common quantum bus in a chiral quantum network. 
	The metamaterial waveguide is placed in a tube which is below the
	superconducting temperature (see experiments in in Ref.~\cite{Magnard2020}). To suppress the thermal microwave noise, each 
	node is placed in a dilution refrigerator which temperature is around 
	$T\sim 10~\text{mk}$.}
	\label{fig5m}
\end{figure}
By considering multiple nodes interacting with the same 
metamaterial waveguide, our proposal in Fig.~\ref{fig1m} can be extended as a chiral quantum 
network. The schematic cascaded network is
depicted in Fig.~\ref{fig5m}, where nodes are placed in separated dilution 
refrigerators with temperature $\sim10~\text{mK}$ to suppress the thermal noise.
As discussed in Ref.~\cite{Magnard2020}, the metamaterial waveguide can be inside a multi-sleeve tube which is below the superconducting critical temperature. This experimental method allows to connect transmons located in different cryogenic refrigerators.
Given that the transmons' frequencies are identical, 
the interaction Hamiltonian can be written as 
\begin{equation}
H_{\text{int},2}=\hbar g_{0}\sum_i\sum_{lk}e^{i\omega_{q}t}\sigma_i^{+}a_{lk}\phi^{*}_{lk}(x_i,t)+\text{H.c.}, \label{Hinto2}
\end{equation} 
where $x_i$ is the coupling position of the $i$th node. As discussed in 
Appendix C, we can derive the cascaded master equation for multiple nodes by tracing 
over the waveguide's freedoms. Taking that two transmons chirally decay/absorb the right propagating photons for example, the evolution is governed by
\begin{eqnarray}
\dot{\rho}_s(t)&=&-iH_{\text{eff}}\rho_s+i\rho_sH^{\dagger}_{\text{eff}}+\mathcal{L}\rho_s\mathcal{L}^{\dagger}, \notag  \\
H_{\text{eff}}&=&-i\sum_i\frac{\Gamma_R}{2}\sigma_i^{+}\sigma_i^{-}-i\sum_{i>j}\Gamma_R\sigma_i^{+}\sigma_j^{-}.
\label{cascademe}
\end{eqnarray}
where $\rho_s$ is the reduced density matrix operator for two transmons, $\Gamma_R$ is the decay rate to the right side,
and $\mathcal{L}=\sqrt{\Gamma_R}(\sigma_1^{-}+\sigma_2^{-})$ is the 
collective jump operator. The last term in $H_{\text{eff}}$ 
is unique to the cascaded quantum system, and describes the irreversible 
process that a photon emitted by transmon $j$ will be reabsorbed by 
transmon 
$i$, while the information back flow is prevented~\cite{Lodahl2017}.
As discussed in Appendix C, when deriving the cascaded master 
Eq.~(\ref{cascademe}), we assume that the propagating time 
$\tau_{ij}=(x_i-x_j)/v_{ln}$ between two nodes is much smaller than the 
decaying time scale $\Gamma_R^{-1}$. Therefore, we can adopt the Markovian 
approximation $t-\tau\simeq t$, and the evolution becomes independent of 
time delay. This approximation is valid only when the separation distance 
$x_{ij}$ is much shorter than the wavepacket's size.
\begin{figure*}[tbph]
	\centering \includegraphics[width=15cm]{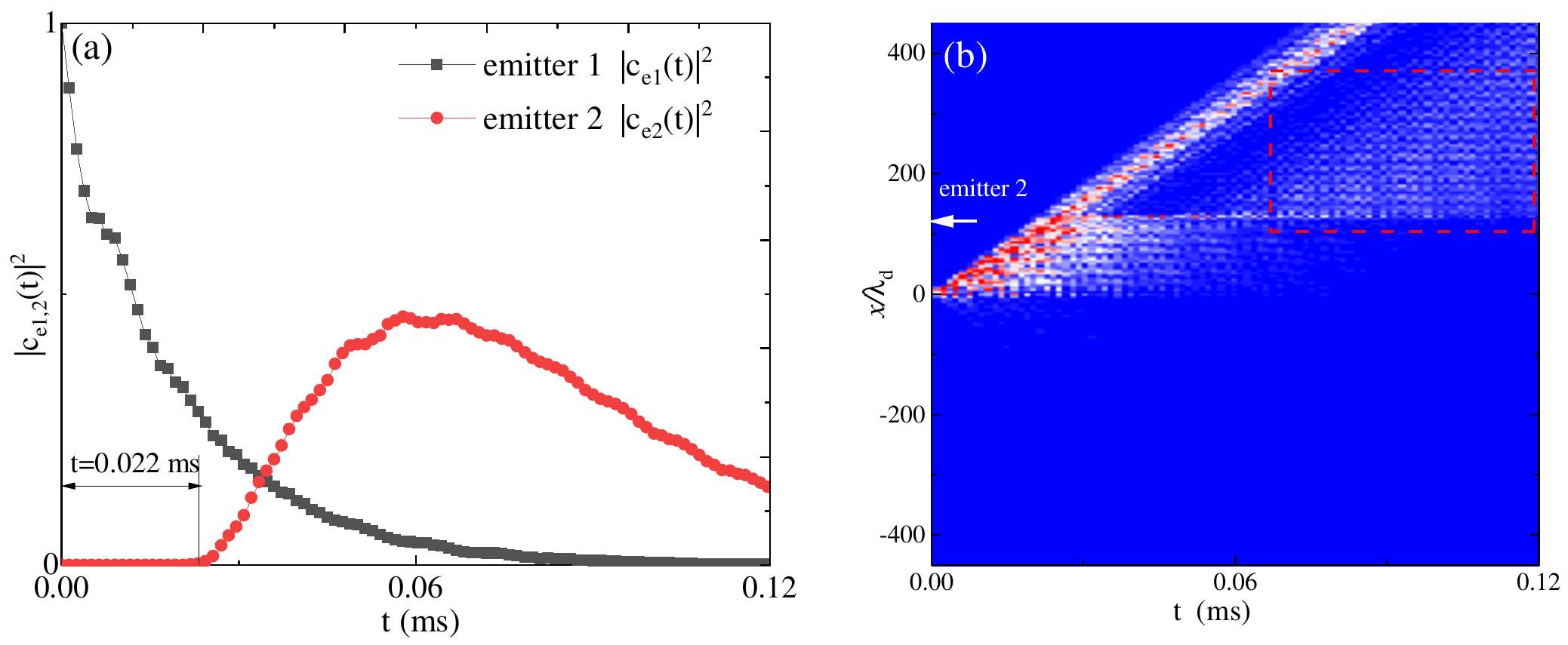}
	\caption{(a) The directional photon flow from transmon 1 ($x=0$) to 
	transmon 2 ($x=125k_d$). Due to the time-delay effect, transmon 2 is 
	excited with a retardation time
	$t=x_{12}/v_0$. (b) The field distribution along the waveguide changes with 
	$t$. At $x=125k_d$, the emitted photon is scattered by transmon 2. Due to 
	directional transport, the scattered field only propagates 
	into the right 
	direction. The adopted parameters are the same with those in 
	Fig.~\ref{fig3m}(c).}
	\label{fig6m}
\end{figure*}

In cascaded master equation~(\ref{cascademe}), the evolution information, such 
as the field distribution and time-decay effects, has been discarded due to taking a 
trace over waveguide's freedoms. To proceed beyond those limitations, our simulation is 
based on the unitary evolution governed by the Hamiltonian in (\ref{Hinto2}), 
which can well describe the time-delay effects~\cite{Wang2022}. Numerical details can be found in Appendix B.
In Fig.~\ref{fig6m}(a), by assuming a single excitation initially 
in transmon 1, we plot the evolution of two transmons with a separation 
distance $x_{2}-x_1=125\lambda_d$. At the frequency adopted in Fig.~\ref{fig3m}(c), the group velocity is around $v_{ln}\simeq 0.83v_0$, which is slower than $v_0$ due 
to the nonlinear dispersion relation around the band edge.
Consequently the wave front arriving at transmon 2 is calculated as
$\tau_{ij}=125\lambda_d/0.83v_0\simeq 0.022\text{ms}$ [see 
Fig.~\ref{fig6m}(a)]. The
field distribution $|\psi_{\gamma}(x,t)|^2$ is depicted in Fig.~\ref{fig6m}(b), 
where one finds that the photon emitted by transmon 1 propagates 
unidirectionally, and the field absorbed by transmon 2 (inside the box) is aslo re-emitted to the right side. Due to no photonic energy back flow, transmon $1$ 
will not be re-excited. Those numerical results show that our proposal can behave as a 
well-performed cascaded quantum network.

\section{Summary and outlooks}
In this work, we propose how to employ the Josephson array as chiral metamaterial waveguide for circuit-QED.
The waveguide is in the form of SQUID chain which impedance is modulated with 
bias currents. When the bias signals are programmed as travelling 
waves, the 
symmetry between the left and right propagating modes is broken due to the
Brillouin-scattering process. 
We also discuss the quantum optical phenomena by considering superconducting qubits coupling to this metamaterial waveguide. By applying the optimized modulating parameters, the qubit will emit photons unidirectionally, and chiral factor can approach 1.
Last we extend our proposal as a multi-node quantum network, and
demonstrate that the chiral transport between remote nodes can be realized.
Compared with routing microwave photons unidirectionally with the bulky 
ferrite circulators, our proposal
does not require strong magnetic field, and the direction can be freely tuned 
by the programmed bias signals. 

Note that we only focus on 
slow travelling waves to modulate the SQUID-chain's impedance. Exploring other 
modulating 
parameters or forms, such as standing-wave modulations and pulses with 
different shapes, might allow us to observe more intriguing QED phenomena in   
this metamaterial platform. 
As discussed in experimental studies in Refs.~\cite{Martinez2019,Planat2020}, 
the SQUID number in a single metamaterial waveguide can be around $10^3\simeq 10^4$, 
indicating that our proposal is within the 
capability of current technology. We hope that our work can inspire more efforts being devoted 
to exploiting SQUID metamaterials for controlling microwave photons in SQC setups.

\section{Acknowledgments}
The quantum dynamical simulations are based on open source code 
QuTiP~\cite{Johansson12qutip,Johansson13qutip}. 
X.W.~is supported by
the National Natural Science
Foundation of China (NSFC) (Grant No.~12174303 and No.~11804270), and China 
Postdoctoral Science Foundation No.~2018M631136. W.X.L. was supported by the 
Natural Science Foundation of Henan
Province (No. 222300420233).
H.R.L. is supported by the National Natural Science
Foundation of China (NSFC) (Grant No.11774284).

\section*{APPENDICES}
\setcounter{equation}{0}
\renewcommand{\theequation}{A\arabic{equation}}
\setcounter{figure}{0}
\renewcommand{\thefigure}{A\arabic{figure}}\

\begin{appendix}	
\section{Deriving the dispersion relation of the modulated SQUID-metamaterial waveguide}
We start from the left side of Eq.~(\ref{diff1}), i.e., the difference terms 
related to time-dependent inductance $L_j(t)$. To 
derive the 
corresponding quasi-continuous differential form, we first rewrite it as
\begin{widetext}
\begin{equation}
	\frac{\phi _j-\phi _{j-1}}{L_{j}(t)}-\frac{\phi _{j+1}-\phi _j}{L_{j+1}(t)}=\frac{\phi _j-\phi _{j-1}}{L_{j}(t)}-\frac{\phi _{j+1}-\phi _{j}}{L_{j}(t)}+\frac{\phi _{j+1}-\phi _{j}}{L_{j}(t)}-\frac{\phi _{j+1}-\phi _j}{L_{j+1}(t)}.
	\label{Lequation}
\end{equation}
\end{widetext}
In this work we assume that each SQUID's size is much smaller than 
wavelengths of both
modulation wave and microwave photons. Therefore, we replace $\phi 
_{j}(t)\rightarrow \phi(x,t)$ and $f(j,t)\rightarrow f(x,t)$. Note that
$G_j(t)$ becomes
\begin{equation}
G(x,t)=\frac{1}{L_{0}}\left[\alpha_{0}+\delta\alpha f(x,t)\right].
\end{equation}
By replacing $x=id_{0}$, Eq.~(\ref{Lequation}) is changed as 
\begin{widetext}
\begin{equation}
	\frac{\phi_j-\phi _{j-1}}{L_{j}(t)}-\frac{\phi _{j+1}-\phi _{j}}{L_{j}(t)}+\frac{\phi 
	_{j+1}-\phi _{j}}{L_{j}(t)}-\frac{\phi _{j+1}-\phi 
	_j}{L_{j+1}(t)}=-G(x,t)\frac{\partial^{2} 
	\phi(x,t)}{\partial x^{2}} d_{0}^{2}-\frac{\partial G(x,t)}{\partial x}\frac{\partial 
	\phi(x,t)}{\partial x} d_{0}^{2}.
\label{leq}
\end{equation}
Similarly, the left side in Eq.~(\ref{diff1}) which contains capacitance terms
can also be written in quasi-continuous differential form. Finally, the 
nonlinear wave function of the modulated SQUID waveguide is derived as
\begin{equation}
	C_g\frac{\partial^{2} \phi(x,t)}{\partial t^{2}}- C_{J}  \frac{\partial^{4} 
	\phi(x,t)}{\partial 
	t^{2}\partial x^{2}}d_{0}^{2}=G(x,t)\frac{\partial^{2} \phi(x,t)}{\partial x^{2}} 
	d_{0}^{2}+\frac{\partial G(x,t)}{\partial x}\frac{\partial 
	\phi(x,t)}{\partial x} d_{0}^{2},
	\label{diff_eqC}
\end{equation}
which is equivalent to the form in Eq.~(\ref{waveeq}).

Given that the modulation is of the travelling wave form, one can decompose the
wave function in Eq.~(\ref{diff1}) as a matrix form by employing the orthogonal 
relations. The dispersion relation is obtained by solving the following 
quadratic eigenvalue problem 
\begin{equation}
	\left[\omega_{l}^{2}(k)\hat{M}_{2}+\omega_{l}(k)\hat{M}_{1}+\hat{M}_{0}\right]\hat{U}(k)=0,
\end{equation}
where $M_{1,2}$ are diagonal matrices and expressed as 
\begin{gather}
	\hat{M}_{2}=\text{diag}\left[..., -c_J(d_{0})^{2}\left( k+nk_{d} \right) ^2-c_g , ... \right], \\
	\hat{M}_{1}=\text{diag}\left[..., -2n\Omega_{d}[c_J(d_{0})^{2}\left( 
	k+nk_{d} \right) ^2+c_g], 
	... \right].
\end{gather}
The matrix $\hat{M}_{0}$ is
	\begin{eqnarray}
		\hat{M}_{0}&=&\left(
		\begin{matrix}
			\ddots & \ddots & \ddots & 0 & 0 & 0 &	\vdots \\
			... & T_{n-1,n-2} & T_{n-1,n-1} & T_{n-1,n} &0 & 0 & ... \\ 
			... & 0 & T_{n,n-1} & T_{n,n} & T_{n,n+1} & 0 & ... \\ 
			... & 0 & 0 & T_{n+1,n} & T_{n+1,n+1} & T_{n+1,n+2} & 0 \\ 
			\vdots & 0 & 0 & 0 & \ddots & \ddots &	\ddots
		\end{matrix}
		\right),
	\end{eqnarray}
	where 
	\begin{gather}
		T_{n,n}=-\left( n\Omega_{d} \right) ^2\left[c_J(d_{0})^{2}\left( k+nk_{d} \right) ^2+c_g\right]+\frac{\alpha_{0}}{l_{0}}\left( k+nk_{d} \right) ^2, \\
		T_{n,n\pm1}=\frac{\delta\alpha}{2l_{0}} \left\{ \left( k+\left( n\pm1 \right) k_{d} \right) ^2+\left( k+\left( n\pm1 \right) k_{d} \right) k_{d} \right\}.
	\end{gather}
\end{widetext}

We find that the Josephson capacitance $c_J$ only appears in the diagonal terms 
of 
$\hat{M}_{0,1,2}$, from which on can easily find that  $c_J$ can be neglected under the following condition
\begin{equation}
	c_J(d_{0})^{2}\left( k+nk_{d} \right) ^2\ll c_g.
	\label{cJcon}
\end{equation}
Further discussions of nonlinear dispersion led by $c_J$
can be found in Sec.~II of the main text.

\setcounter{equation}{0}
\renewcommand{\theequation}{B\arabic{equation}}

\section{Numerical methods for simulating spontaneous emission and 
non-Markovian dynamics}
Although the master equation can well describe the spontaneous emission of emitters,
the Born-Markovian approximation is not valid when the interaction 
strength is comparable to the band width of baths. Moreover, the information of 
emitted photons will be traced off when deriving the master equation. To avoid 
those, our simulation is based on the unitary evolution governed by the 
original Hamiltonian in Eqs.~(\ref{Hinto}, \ref{Hinto2}). In the single-excitation 
subspace and taking for the case of two transmons for example, 
the system's state is written as 
$$|\psi(t)\rangle=\sum_{lk} 
c_{lk}(t)|g,g,1_{lk}\rangle+c_{e1}(t)|e,g,0\rangle+c_{e2}(t)|g,e,0\rangle,$$  
where $|1_{lk}\rangle$ represents the single excitation being in $lk$th mode, 
and $|0\rangle$ corresponds to the waveguide in its vacuum.
Since the Hamiltonian is expressed in momentum space, we first need to 
calculate waveguide's eigen-frequencies and wavefunctions by employing the 
method presented in Appendix A. Next we discretise modes in the first BZ $k\in (-0.5k_m, 
0.5k_m]$ with a large number $N$, which is equal to considering a waveguide 
with a
length $L=N\lambda_m$. A similar method can be found in Ref~\cite{Wang2022}. 
In the 
single-excitation subspace, the Hamiltonian in Eq.~(\ref{Hinto}) 
[Eq.~(\ref{Hinto2})] can be mapped into a matrix with dimension $N+1$ ($N+2$) 
when the system contains a single emitter (two emitters). 
Taking two emitters ($Q=2$) for example, the corresponding matrix is written as
\begin{widetext}
	\begin{eqnarray}
		H_{\text{int}}=\left[ \begin{matrix}
			w_{lk_1}&		0&		...&		0&		g_0\phi_{lk_1}(x_1,t) 
			&		g_0\phi_{lk_1}(x_2,t)\\
			0&		w_{lk_2}&		\ddots&		...&		g_0\phi_{lk_2}(x_1,t) &		g_0\phi_{lk_2}(x_2,t)\\
			\vdots&		\ddots&		...&		0&		\vdots &		\vdots \\
			0&		...&		0&		w_{lk_N}&		g_0\phi_{lk_N}(x_1,t) &		g_0\phi_{lk_N}(x_2,t) \\
			g_0\phi^*_{lk_1}(x_1,t) &		g_0\phi^*_{lk_2}(x_1,t)&		...&	g_0\phi^*_{lk_N}(x_1,t)&		w_{qa} &	0\\
		g_0\phi^*_{lk_1}(x_2,t)&		g_0\phi^*_{lk_2}(x_2,t)&		
		...&		g_0\phi^*_{lk_N}(x_2,t)&		0 &		w_{qb}\\
		\end{matrix} \right],
		\label{Hintmat}
	\end{eqnarray}
\end{widetext}
where $\phi_{lk_i}(x_j,t)$ is numerically obtained via methods in Appendix A. 
After obtaining the matrix form in Eq.~(\ref{Hintmat}), we can numerically 
solve the evolution governed by the Schr\"{o}dinger equation. 
At certain time $t=t_i$, the amplitude $c_{kl}(t)$ of each mode is recorded, 
and the field distribution can be recovered by substituting them into 
Eq.~(\ref{field_D}). Employing this method, we obtained both transmon's and 
waveguide's evolution shown in the main text.

\setcounter{equation}{0}
\renewcommand{\theequation}{C\arabic{equation}}

\section{Cascaded master equation for multi-nodes system}
In this part we will derive the cascaded master equation for multiple emitters mediated 
by the metamaterial waveguide.
By expanding the Schr\"{o}dinger equation to the second order, the evolution of 
the system is expressed as~\cite{Scully1997} 
\begin{eqnarray}
\dot{\rho}_s(t)&=&-\frac{i}{\hbar}\mathrm{Tr}_R{\left[ H_{\mathrm{int},2}(t),\rho (t) 
\right]}- \\
&&\!\frac{1}{\hbar ^2}\mathrm{Tr}_R\int_{t_0}^t{\left[ H_{\mathrm{int},2}(t),\left[ H_{\mathrm{int},2}(t^\prime),\rho (t^\prime) \right] \right]}dt^\prime,\label{Hrho}
\end{eqnarray}
where $\rho$ ($\rho_s$) is the density matrix operator of the whole system (the 
emitters), and
$\text{Tr}_R$ represents taking a trace over the waveguide's freedoms. 
We assume that the waveguide is always approximately in its vacuum state. 
Therefore, the 
correlation relations for the waveguide's modes satisfy $\langle a_{lk} \rangle=\langle a^\dagger_{lk} 
\rangle=\langle a^\dagger_{lk}  a_{lk}  \rangle=0$ and $\langle a_{lk}  
a^\dagger_{lk}  \rangle=1$. By substituting 
those relations into Eq.~(\ref{Hrho}), we obtain 
\begin{eqnarray}
\dot{\rho}_s(t)&=&-g_{0}^{2}{\frac{L}{2\pi}}\sum_{i,j}\int_{t_0}^t dt' A\left( 
x_i,x_j,t,t^{\prime} \right)  \notag \\
&& \big[ \sigma _{i}^{+}\sigma _{j}^{-}\rho_s (t^\prime) -\sigma _{j}^{-}\rho_s (t^\prime)\sigma _{i}^{+} \big] +\mathrm{H}.\mathrm{c}., \label{rhot}
\end{eqnarray}
where the correlation function $A\left( x_i,x_j,t,t^{\prime} \right)$ is 
defined as~\cite{Calaj2019}
\begin{eqnarray}
&&A\left( x_i,x_j,t,t^{\prime} \right)=e^{i\omega _q\left( t-t^{\prime} \right)}\sum_{lk}\phi _{lk}^{*}(x_i,t)\phi _{lk}(x_j,t) \notag \\
&=&\sum_{ln}|u_{ln}(k_{ln})|^2
\int_{-\infty}^{\infty}{d}(\delta k)e^{-i\delta kv_{ln}\left[ (t-t^\prime)-\frac{ x_{ij} }{v_{ln}} \right]} \notag \\
&=&\sum_{ln}\frac{2\pi}{|v_{ln}|} |u_{ln}(k_{ln})|^2
\delta \left[(t-t^\prime)-\frac{x_{ij} }{v_{ln}} \right],
\end{eqnarray}
with $x_{ij}=x_i-x_j$ being the distance between two emitters. We have 
neglected the phase terms $e^{i\left( k_{ln}+k_d \right) \left( x_i-x_j 
\right)}$ in $A\left( x_i,x_j,t,t^{\prime} \right)$. Note that the integral 
regime is within $t^\prime\in[t_0,t]$. Therefore, the $\delta$-funtion will 
produce non-zero value in Eq.~(\ref{rhot}) only under the condition 
$t-x_{ij}/v_{ln}\leq t$. The delay-time $\tau_{ij}=x_{ij}/v_{ln}$ ($i\neq j$) 
corresponds to the propagating time between two separated emitters. When 
$x_i>x_j$ 
($x_i<x_j$), the non-zero time-delay correlation is mediated by the right 
(left) propagating modes with $v_{ln}>0$ ($v_{ln}<0$). 
Given that $\tau_{ij}$ is much smaller than the time scale of spontaneous 
emission, we can replace $t-\tau\rightarrow t$, and Eq.~(\ref{rhot}) is 
simplified as 
\begin{eqnarray}
	\dot{\rho}_s(t)&=&\Big\{\frac{\Gamma_R+\Gamma_L}{2} \sum_{i}\big[ \sigma _{i}^{+}\sigma _{i}^{-}\rho_s (t) -\sigma _{i}^{-}\rho_s (t)\sigma _{i}^{+} \big] \notag \\
	&&-\sum_{i> j} \Gamma_R \big[ \sigma _{i}^{+}\sigma _{j}^{-}\rho_s(t) -\sigma _{j}^{-}\rho_s(t)\sigma _{i}^{+} \big]+ \notag \\
	&&\!\sum_{i<j} \Gamma_L \big[ \sigma _{i}^{+}\sigma _{j}^{-}\rho_s(t) -\sigma _{j}^{-}\rho_s(t)\sigma _{i}^{+} \big]\Big\}\!	
	 +\!\mathrm{H}.\mathrm{c}., 
	\label{rhot1}
\end{eqnarray}
where $\Gamma_{R(L)}$ is the decay rate into right/left side, which are expressed in Eq.~(\ref{GammaRL}). In Eq.~(\ref{rhot1}) 
we have employed the properties of $\delta$-function 
\begin{eqnarray}
\int_{t_0}^t \delta \left( t-t^\prime-\frac{x_{ij} }{v_{ln}} \right)\rho_s (t^\prime)dt'=\left\{
\begin{array}{lr}
	\frac{1}{2}\rho_s (t) \quad x_{ij}=0,   \\
	\rho_s (t)  \quad x_{ij}/v_{ln}=0^+, \\
	0 \quad x_{ij}/v_{ln}<0. \notag 
\end{array} 
\right.
\end{eqnarray}
When the chiral factor of the whole system approaches $\beta_\pm\simeq 1$, we
obtain the cascaded master equation in the main text. 
\end{appendix}

%

\end{document}